\documentclass[conference]{IEEEtran}
\IEEEoverridecommandlockouts
\usepackage{cite}
\usepackage{amsmath,amssymb,amsfonts}
\usepackage{algorithmic}
\usepackage{graphicx}
\usepackage{caption}
\usepackage{subcaption}
\usepackage{textcomp}
\usepackage[graphicx]{realboxes}
\usepackage[table,xcdraw]{xcolor}
\def\BibTeX{{\rm B\kern-.05em{\sc i\kern-.025em b}\kern-.08em
    T\kern-.1667em\lower.7ex\hbox{E}\kern-.125emX}}
\IEEEaftertitletext{\vspace{-2\baselineskip}}
\begin{document}

\newcolumntype{C}[1]{>{\centering\arraybackslash}p{#1}}
\title{Efficient and Secure Energy Trading with Electric Vehicles and Distributed Ledger Technology}

\author{
\IEEEauthorblockN{Conor Mullaney}
\IEEEauthorblockA{\textit{} 
\textit{Toshiba Europe Ltd.}\\
Bristol, United Kingdom \\
conor.mullaney@toshiba-bril.com}

\and
\IEEEauthorblockN{Adnan Aijaz}
\IEEEauthorblockA{\textit{} 
\textit{Toshiba Europe Ltd.}\\
Bristol, United Kingdom \\
adnan.aijaz@toshiba-bril.com}
\and
\IEEEauthorblockN{Rasheed Hussain}
\IEEEauthorblockA{\textit{} 
\textit{University of Bristol}\\
Bristol, United Kingdom \\
rasheed.hussain@bristol.ac.uk}
}

\maketitle

\begin{abstract}
Efficient energy management of Distributed Renewable Energy Resources (DRER) enables a more sustainable and efficient energy ecosystem. Therefore, we propose a holistic Energy Management System (EMS), utilising the computational and energy storage capabilities of nearby Electric Vehicles (EVs), providing a low-latency and efficient management platform for DRER. Through leveraging the inherent, immutable features of Distributed Ledger Technology (DLT) and smart contracts, we create a secure management environment, facilitating interactions between multiple EVs and energy resources. Using a privacy-preserving load forecasting method powered by Vehicular Fog Computing (VFC), we integrate the computational resources of the EVs. Using DLT and our forecasting framework, we accommodate efficient management algorithms in a secure and low-latency manner enabling greater utilisation of the energy storage resources. Finally, we assess our proposed EMS in terms of monetary and energy utility metrics, establishing the increased benefits of multiple interacting EVs and load forecasting. Through the proposed system, we have established the potential of our framework to create a more sustainable and efficient energy ecosystem whilst providing measurable benefits to participating agents.
\end{abstract}

\section{Introduction}
The growing adoption of Distributed Renewable Energy Resources (DRERs) in the energy industry facilitates the decentralisation of current energy markets, enabling consumers to become Prosumers \cite{ZHANG20181}. Additionally, if DRER are built on suitable frameworks, they reduce single points of failure weaknesses increasing energy reliability \cite{IOTAP2P}. However, the most abundant and accessible DRER such as solar and wind power are difficult to manage. They are intermittent and dependant on uncontrollable factors such as weather, requiring flexible and low-latency Energy Management Systems (EMS) to accommodate their variable generation \cite{Safdar2013AMicro-grids}.

Electric Vehicles (EVs) remain an underutilised resource, spending approximately 96\% of their time idle, leaving their computational and ample energy storage resources to remain unused \cite{kiaee2015estimation}. This has led to research paradigms such as Vehicular Fog Computing (VFC) and Vehicle-to-Grid (V2G), to utilise the computational and energy storage resources of these idle EVs \cite{hu2017survey,ouramdane2022home}.

The integration of EVs as energy management resources through their computational and energy storage capabilities offers a promising solution for effectively managing and maximising the potential of DRER. However, managing the interactions between EVs and DRER through IoT devices presents security, privacy, and resource allocation challenges \cite{parikh2019security,ravi2022utilization}. Inefficient management of these challenges can lead to wasted resources and privacy breaches, making the system unfeasible. 


In summary, with the right system in place, we can realise the full potential of EVs as energy management resources utilising their computational and energy storage resources, creating a more sustainable and efficient energy ecosystem. Nevertheless, in order for this to be effective, the EMS needs to provide a secure platform,  low-latency response, and economically feasible incentives to be viable. To fill these gaps, the main contributions of this work are as follows:

\begin{itemize}
    \item We present a cohesive EMS that integrates the computational and energy storage capabilities of EVs in conjunction, enabling effective management of distributed and renewable energy resources.
    
    \item We develop an energy trading system that exploits the key benefits of \emph{IOTA Tangle} and \emph{IOTA smart contract protocol (ISCP) framework}. The proposed system underpins the security aspects of our system, allowing multiple EVs and Prosumers to interact securely. 

    \item We utilise a privacy-preserving federated learning method for load forecasting, enabling EVs to seamlessly join and exit our trading system. Additionally, by forecasting energy production, we enable efficient matching algorithms to maximise the potential of our EV energy storage resources.  
    
    \item We provide simulations-based results to assess the proposed system in various trading scenarios using empirical data. Our results focus on the effect of our system in terms of monetary benefit and energy utility.
    
\end{itemize}
The rest of this paper is organized as follows. Section \ref{sec:relatedwork} covers related work and preliminaries on the subject matter. Section \ref{sec:background} discusses background and Section \ref{sec:proposed} describes our proposed EMS. We discuss experimental setup in Section \ref{sec:experiments}  and results in \ref{sec:results}. The paper is concluded in section \ref{sec:conclusions} with some directions for future work.

\section{Related Work}
\label{sec:relatedwork}
This section covers the related work surrounding using EVs for the energy management of DRER. Given that there are two primary EV resources of computation and energy storage, research into IoT-EV integration often explores these factors separately. However, by considering these capabilities in conjunction, we aim to provide a holistic EV energy management solution.
\subsubsection{V2G}
Integrating EV energy storage resources and the grid is an extensive area of research relevant to our study, encompassing both Vehicle to Home (V2H) and Vehicle to Grid (V2G) \cite{vadi2019review}. This research area highlights the potential of using EV batteries as energy storage resources as a promising management solution for DRER. In \cite{ouramdane2022home}, the authors present the potential of V2H, suggesting an EV acting as a backup should participate in energy management for a home, providing monetary benefits to the residents. The proposed system defines clear modes of operation based on power levels in the system, only considering energy forecasting as a future research direction that would allow more adaptive protocols. In addition, \cite{ravi2022utilization} discusses the potential of an EV for local utilisation as an energy storage unit for the home, or to help congestion within the grid. The authors suggest that a Virtual Power Plant takes the role of aggregating the data from the grid and using it to forecast energy management; however, this introduces a layer of trust reducing the privacy of the system. Though V2H literature provides promising solutions for integrating EVs, they do not consider larger-scale numbers of EVs, and the effect of efficient algorithms utilising forecasts to benefit participating agents.

V2G literature scales the system to multiple EVs and Prosumers. In \cite{real-timeEMMicro}, the authors discuss the potential of multiple EVs as movable energy storage to manage a microgrid. They establish monetary benefit as a key metric for performance and do not explore the potential benefit of forecasted algorithms. The authors instead focus on a real-time management algorithm to reduce security risks as no data needs to be stored. Alternatively, \cite{xu2022short} suggests a matching algorithm to be used after load forecasting and before vehicle integration. The authors suggest a priority mechanism that increases the grid's quick-response capability but do not address the security concerns of the surrounding multiple interacting IoT devices and assume load forecasting has been implemented.

\subsubsection{VFC}
Vehicular Fog Computing (VFC) extends the Fog Computing (FC) paradigm to use available computational resources (in this case EVs) near IoT devices to perform relevant computational tasks. FC provides an edge computing solution with faster execution than alternatives, such as cloud computing. This is at the expense of higher energy consumption and limited scalability, as our resources are limited to those close to our IoT devices \cite{yi2015survey}. This type of infrastructure works best for real-time applications needing low-latency computation to ensure faster response times, such as energy management.

Jaiswal \emph{et al.} \cite{jaiswal2021fog} highlight the potential of Fog resources, exploring the efficacy of FC energy prediction in smart grids. They explore Long Short Term Memory (LSTM) neural networks for energy forecasting over alternatives such as Recurrent Neural Networks, due to their resistance to vanishing and exploding gradients. However, they do not consider the privacy aspects surrounding sharing sensitive data needed for energy forecasting.
 Similarly, Zhang \emph{et al.}  \cite{zhang2022federated} extend this step, presenting LSTM load prediction in a federated learning context. The privacy advantages of federated learning are discussed, as raw data does not need to be shared amongst all participants. However, they do not consider if nodes are unable to carry out LSTM learning due to constrained devices. Therefore, we propose that EVs implement federated load prediction using an LSTM, implementing protocols to handle moving EVs.
 
\section{Background and Preliminaries}
\label{sec:background}
\subsection{Physical Infrastructure}
The authors of \cite{hou2016vehicular,ning2019vehicular} discuss the relevant infrastructure needed for VFC. The authors define critical scenarios for VFC-specific FC implementation, identifying a reliable communication network as the backbone of a VFC network. Wireless communication mediums called Road-Side-Units (RSU) are discussed as a potential solution, enabling reliable and fast-acting communication and protocols.

Low latency communication is also prevalent in V2G literature. Different existing works, e.g., \cite{ravi2022utilization} explore V2G as a real-time system that needs real-time monitoring and low-latency response to deal with the intermittency of RERs and EVs' battery State of Health (SoH). Therefore, similar to VFC, a robust communication structure is required that is able to handle the real-time transfer of data, such as traveling conditions, e.g. weather and battery SoH. 

Not only communication needs, but there are also grid requirements. Both DRER, and EVs pose challenges to the current centralised power network with its unidirectional power flows, needing stable energy production methods to function efficiently \cite{Weiss2019AGenerators}. A promising solution in literature is the concept of a smart grid which has increased in adoption around the world with projects such as GridWise \cite{melton2013gridwise}, and the Gotland smart grid \cite{wallnerstrom2018analysis}. Smart grids are distributed energy networks that can effectively manage distributed energy resources. They achieve this through two-way communication features, sophisticated computing systems, and sustained power flows between intelligent components \cite{Smart_Grids}.

\begin{figure}[h]
    \vspace{-1em}
    \centering
    \includegraphics[width=0.48\textwidth]{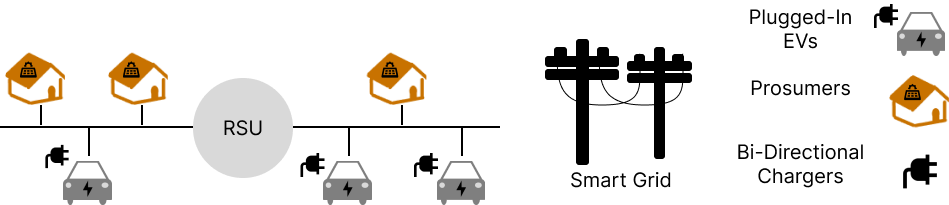}
    \caption{Infrastructure layer.}
    \label{PhysicalLayer}
    \vspace{-1em}
\end{figure}

Therefore, we propose an infrastructure represented in Fig. \ref{PhysicalLayer}, providing bi-directional energy capabilities and a distributed framework allowing EVs to be integrated as energy management resources. 

In our smart grid scenario, we assume that the bi-directional chargers can accommodate all available energy exchange between the system's entities, leading to more optimal use of resources. The aforementioned RSU's, act as a communication medium to connect the cyberstructural and physical infrastructure allowing the interacting agents to utilise the functionaility on the cyberstructural layer.

\subsection{Cyberstructure}
 DLT is a well-known technology for enabling trustless interactions between entities in energy markets \cite{Wang2019}.  DLT provides a decentralised trading platform whilst also providing benefits in terms of privacy, cost, and security \cite{Shuaib2018}. The potential of DLT as a secure management solution is agreed upon in VFC and V2G literature \cite{kong2021secure,iqbal2021secure}. A blockchain-based resource management scheme for VFC is explored in \cite{kong2021secure}. Here, DLT is proposed as a solution to the security and privacy risks of FC networks due to inherent beneficial features of decentralisation and trustless exchange. The use of smart contracts to mediate agreements between fog nodes and IoT devices is also explored as a potential access management and authorisation solution. Moreover, V2G DLT implementation is explored in \cite{iqbal2021secure}, which uses DLT blockchain technology as a security management framework for an energy trading model. It underscores the potential of smart contracts in trustless environment. 
 
Not all DLTs are suited for all tasks; \cite{wen2020blockchain} comprehensively reviews various DLTs in an IoT-Centric use case, detailing the advantages and disadvantages of specific architectures and consensus protocols. The authors identified the Tangle as a suitable Directed Acyclic Graph (DAG) based DLT for IoT environments. The Tangle is a DLT built by the IOTA foundation aimed at an IoT environment \cite{Popov2018TheTangle}. It has a decentralised, secure infrastructure without needing high computational power to function efficiently. At the same time, it has high scalability, high throughput, and low-latency \cite{Salimitari2018ANetworks}. 

The benefits of IOTA have been identified in past literature, e.g., \cite{IOTAV2G} suggests that using the IOTA Tangle is a promising solution for trustless exchange in V2G networks. However, they do not consider ISCP since it was not available at the time. On the other hand, \cite{IOTAP2P} explores IOTA, implementing ISCP for P2P energy trading between microgrids, highlighting the potential of the sharded ISCP architecture with privacy and scalability advantages over alternative DLT options whilst maintaining security. ISCP aims for a completely decentralised IOTA ecosystem, but currently still utilises a centralised entity known as a coordinator, a security weakness to the network. Despite this, the potential of ISCP and planned decentralisation in IOTA 2.0 make it a promising option for our research.

\begin{figure}[ht]
    \vspace{-1em}
    \centering
    \includegraphics[width=0.48\textwidth]{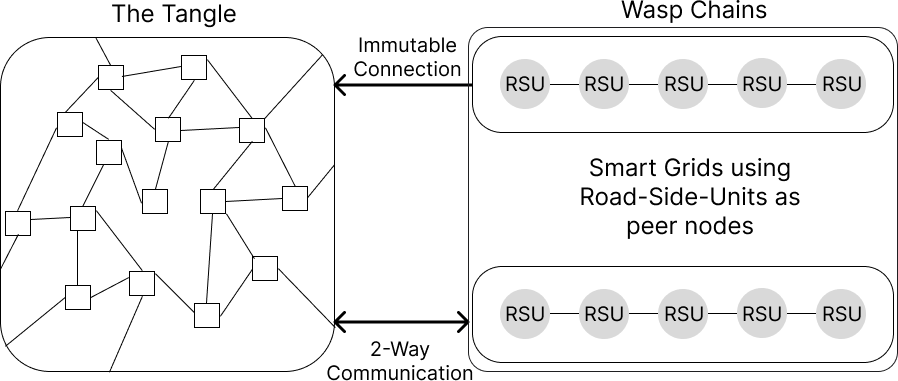}
    \caption{Sharded blockchains connecting to the Tangle.}
    \label{Shards}
    \vspace{-1em}
\end{figure}

To utilise sharded ISCP architecture, each smart grid wanting to integrate EVs as energy management resources runs its own ``Wasp chain". These chains use RSU's as peer nodes and have the relevant energy management smart contracts deployed upon them as shown in Fig. \ref{Shards}. Using the consortium nature of our Wasp chains allows us to take advantage of a public-private DLT split increasing security and privacy as we can control who has access to the Wasp chain. This can be achieved with an appropriate DLT-based authentication scheme on our sharded blockchains, such as presented in \cite{singla2018blockchain}, underpinning CA authentication with DLT. This enables secure and private interactions between multiple EVs and Prosumers leveraging, the scalable ISCP architecture to its full advantage.

\begin{figure}[h]
    \centering
    \includegraphics[width=0.48\textwidth]{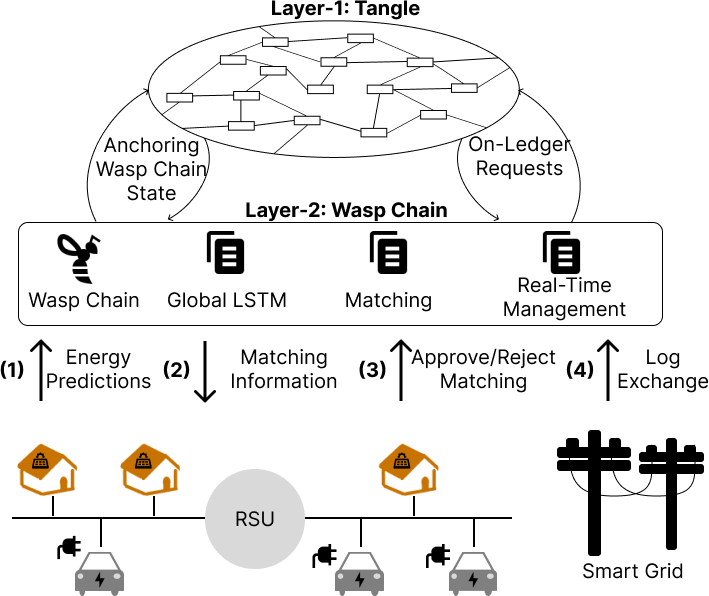}
    \caption{System overview.}
    \label{overview}
\end{figure}

\section{EV energy trading framework}
\label{sec:proposed}
By taking into account the physical infrastructure needs and the cyberstructural needs we have developed a Cyber-Physical System architecture displayed in Fig. \ref{overview}.

The smart grid represents the infrastructure layer and the interacting agents in the network. At the infrastructure layer, all required information is provided between matched entities including energy needs, battery capacity, matching ID, CA tokens, etc. The Wasp chain layer is layer-2 of the Tangle upon which the smart contracts communicate and reach consensus. The matching contract handles energy bids from the agents. The real-time management contract logs energy exchange for matchings. Then, the LSTM contract handles the aggregation and distribution of the LSTM parameters in federated learning. Finally, the Tangle DLT anchores the state of these smart contracts in the form of a hash value to maintain immutability. This occurs when the smart contracts experience a change of state. The changes in state only have to be verified by participating agents in the consortium Wasp chain. This, in conjunction with an authentication scheme, as described, allows us to control what information is stored on the Tangle to maintain security whilst keeping privacy of the agents.

\subsection{Load Forecasting}
In traditional federated learning, each agent in the system uses their data to train an LSTM before passing the LSTM parameters to a cloud node. This cloud node then collates the parameters of various models and creates an updated version of the model itself. Each agent then uses this new model to continue training their data \cite{zhang2022federated}.

\begin{figure}[ht]
    \vspace{-0.9em}
    \centering
    \includegraphics[width=0.4\textwidth]{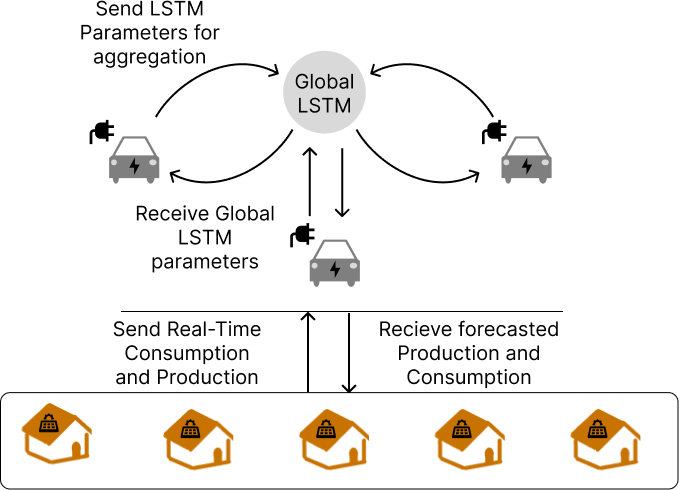}
    \caption{Federated overview.}
    \label{FogAC}
    \vspace{-1em}
\end{figure}

To implement load forecasting through federated learning and VFC, we assume that a Prosumer participating in energy trading has an EV implementing our proposed load forecasting behaviour and the forecasts are used in the trading simulation. An overview of the implemented federated-LSTM is shown in Fig. \ref{FogAC}. Each Prosumer uses an EV acting as a fog node to train a local model of our system. This maintains privacy concerns of the Prosumers as they only need to share the real-time energy production and consumption metrics with the EV node training their LSTM. 

Considering the constraints of the federated paradigm, we assume that each smart grid has its own ``Global" LSTM parameters from which EV fog nodes pull and train their local LSTMs. This provides another benefit as when an EV enters the network to train and predict, they are able to use the ``Global" LSTM model as a starting point. As such, we do not use a traditional federated aggregation method and instead aggregate the LSTM models periodically. This allows new EVs entering the system to have a relatively updated LSTM model from which to make predictions.

\subsection{Matching Algorithms}
Before energy trading occurs, participating agents place energy bids. Prosumers place net energy production/consumption bids and the EVs place available energy and battery capacity bids. This allows us to use matching theory, a common job allocation method in VFC and V2G literature \cite{jia2018double,xu2022short}.

In order to enhance security while facilitating interactions between multiple EVs and multiple Prosumers, we employ a strategy that restricts each agent to a single matching per period and model our EVs as passive chargers. This limitation effectively minimizes the risk of exposure to malicious entities, thereby bolstering security, and resembles the Linear Sum Assignment Problem (LSAP). LSAP aims to find the optimal solution for matching multiple resources (EV resources) to multiple tasks (DRER energy management) based on specific constraints.

To show the benefit of load forecasting in the proposed solution, we display two options. We initially explore a ``Greedy" implementation for the matching strategy, using a First-In-First-Out (FIFO) matching algorithm to match resource providers with requests. This approach offers fast response times and operates in {\it O(n)} time complexity, making it a low-latency matching option that does not need load forecasting. However, there is a risk the algorithm matches the first available resource with the first requester, without considering whether a better match may be available later. Therefore, we consider the Hungarian algorithm, which is a long standing approach in literature \cite{mills2007dynamic}. 

The Hungarian algorithm creates a cost or benefit matrix associated with assigning the requested resources (e.g. energy or capacity) to buyers and sellers and uses the matrix to find the matching with minimum cost. In the proposed algorithm, this cost for producers is the amount of energy not able to be stored in the matched EV, and for net consumers, this was calculated as the unfulfilled energy demand. While the Hungarian algorithm provides an optimal solution for the assignment problem in terms of minimising the total cost, its time complexity of {\it O($n^3$)} may lead to significant overhead and computation in the system, and therefore is only viable considering the load forecasting behaviours of our system.

\section{Experimental Setup}
\label{sec:experiments}
\subsection{Agents}
\begin{figure}[ht]
    \vspace{-0.5em}
    \centering
    \includegraphics[width=0.45\textwidth]{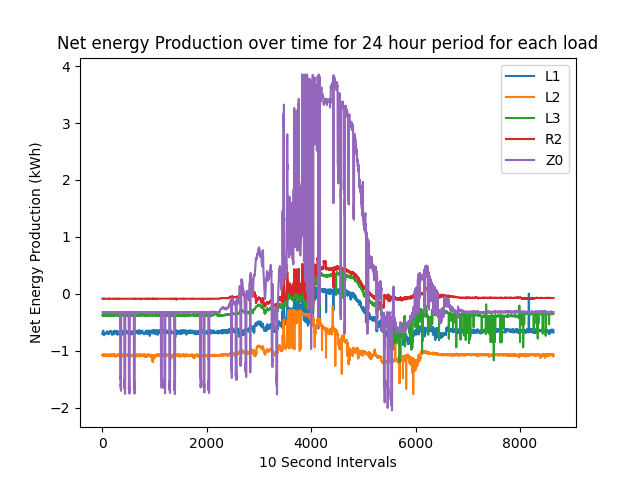}
    \caption{Load profile of each agent over our 24 hour testing period.}
    \label{LoadProfile}
    \vspace{-1em}
\end{figure}

In this study, the loads are simulated based on the data set presented in \cite{Gomes2019-xv}. The data set includes five PhotoVoltaic (PV) generation agents, providing their power production and consumption measurements every 10 seconds over a week period. This data has high temporal resolution, and has been used in past literature to model DRER \cite{IOTAP2P, Gomes2019-xv}. The net production of energy for each Prosumer for the testing period is represented in Fig. \ref{LoadProfile}, demonstrating the general trends throughout the testing period.

The EVs are modeled from an open source python tool ``emobpy" which generates time series data for our EVs \cite{gaete2021open}. This data allows us to model the time series data of the cars in both location, energy consumption, and charging behaviours along with the relevant SoC information at 15-minute intervals. Using this open source tool, we generated time series data for a Tesla Model 3 Long Range AWD, as it is one of the most popular EV models. 
 We use the long-range model, which has a greater battery capacity ($79.5 kWh = C_{T}$). We limited the operable battery $C_{o}$ to minimise battery degradation \cite{kostopoulos2020real}, and reserved a further weeks worth of energy consumption for the EV $E_{r}$ to account for the passiveness of the EVs when charging. This is represented by Equation \ref{capacity_range}.

\begin{equation}
\label{capacity_range}
Range_{cap} = (0.2 * C_{T}) +  E_{r} \leq C_{o} \leq 0.8 * C_{T}
\end{equation}

\subsection{Pricing}
\subsubsection{Pricing}
Considering the buying and selling of energy between EVs and Prosumers, we limit our pricing mechanism by the buying and selling prices to and from the main grid. This ensures that participants in our system will achieve no worse than if they had traded energy with the main grid. The price of buying energy from the main grid $P_{gb}$ is determined to be $29.49p/kWh$ \cite{GridBuyPrice}, and the selling price of energy to the main grid $P_{gs}$ is determined as $6.4p/kWh$ \cite{GridSellPrice} respectively.

When a Prosumer wants to sell excess energy to an EV, the selling price is determined to be the split difference between the buying and selling price to the grid and is shown in Equation \ref{split}. An EV can only sell energy to a Prosumer if the EV has been participating in the network, which can be verified in the DLT. Once the participation has been verified, the price of energy in the EV is determined and sold with a fixed tariff of 10\% once again limited by the main grid prices. This ensures that even if EVs do not keep the energy, they still achieve a 10\% profit on every Wh of energy stored in their batteries. 

 \begin{equation} 
    P_{split} = \frac{P_{gb} + P_{gs}}{2} 
    \label{split}
\end{equation}


\subsection{Solo Testing Environment}\label{SolotestSetup}
Solo is a testing framework that allows us to validate and interact with the smart contracts and inter-chain protocols without requiring their deployment to the Tangle. As such the solo testing environment was used to evaluate the functionality of smart contracts on the Wasp chain, with each entity being able to interact and place bids to smart contracts through Solo Contexts. A full explanation of the solo testing environment can be found on IOTA wiki \cite{iotawiki_2023}. When an energy exchange is successfully achieved, we print all relevant information of the trade to an ``exchange.csv" file: time step, energy exchanged, available energy, battery capacity, matching id, Prosumer and EV, from which we are able to derive our results.

\subsection{Federated-LSTM Setup} \label{federatedTestSetup}
The LSTM setup is implemented using the Python PyTorch library and neural network modules \cite{Pytorch}. Since the implemented LSTM is primarily a proof of concept to demonstrate the potential of using EV computational resources, we did not prioritise a thorough analysis of optimizing the model performance. Rather, our goal was to show that the LSTM could produce meaningful results for the chosen EMS using a reasonably straightforward configuration.

Each LSTM was configured with two layers, each consisting of 32 neurons. The inputs to the model included time, consumption, and production data for the preceding three matching periods, while the outputs included consumption and production predictions for the upcoming matching period. This sliding window approach reduces our system's reliance on storing large amounts of data related to energy production and consumption, which could pose a potential vulnerability. The model was trained using Mean Squared Error (MSE) as the loss function and optimised using the Adam optimiser. Each matching period was 60 seconds whilst the actual trading of energy was at the granularity of the load data set.

\section{Results and Discussion}
\label{sec:results}
Fig. \ref{LoadProfile} displays the load profiles of each Prosumer in our study. As expected for PV Prosumers, all load profiles exhibit peak net energy production during midday (interval 4320) and lowest net energy production at the tail ends of the 24-hour period. Prosumer Z0 demonstrated the highest energy production among all participants, producing approximately three times more energy than the next closest Prosumer R2 during peak production. On the other hand, Prosumer L2 had the highest consumption and did not generate any energy within the 24-hour testing period.

Processing the information available from the ``exchange.csv" file we can evaluate the effectiveness of our system. Table \ref{v2hBenchmark} establishes a baseline of systems common in V2H literature where a single trusted EV is used to smooth a renewable energy Prosumer in an isolated trading format \cite{ouramdane2022home}. There is a significant benefit to all agents involved as they are able to use the energy they have stored in periods of high production and use them in periods of high consumption, experiencing benefits between 8.35-200.42 kWh. Prosumer L2 experiences very little benefit as it produced 0kWh throughout the 24-hour window as shown in Fig. \ref{LoadProfile} and therefore was only able to use the energy available in the EV, 8.35kWh. On the other hand, Z0 experienced the most significant benefit of 200.42 kWh as it was able to store the most energy in the EV for later use, an expected outcome considering its profile as the largest energy producer.

\begin{center}
    \begin{table}[h]
    \vspace{-1em}
    \centering
    \caption{Table showing the benefit of each agent trading with a single EV for a 24-hour window, measured in terms of reduced energy exchange with the main grid.}
    \begin{tabular}{|c|C{0.1\textwidth}|C{0.1\textwidth}|C{0.1\textwidth}|}
    \hline
    \textbf{Agent} & \textbf{Original Grid Exchange (kWh)} & \textbf{Isolated Grid Exchange (kWh)} & \textbf{Absolute benefit (kWh)} \\ \hline
    L1    & 4884.36 & 4825.60 & 58.76   \\ \hline
    L2    & 8672.30  & 8663.95  & 8.35  \\ \hline
    L3    & 2967.63 & 2874.31  & 93.32  \\ \hline
    R2    & 1039.07    & 944.96 & 94.11 \\ \hline
    Z0    & 6558.82  & 6358.40 & 200.42 \\ \hline
    \end{tabular}
    \label{v2hBenchmark}
    \end{table}
    \vspace{-2.5em}
\end{center}

\begin{figure}[h]
    \vspace{-1.5em}
    \centering
    \includegraphics[width=0.48\textwidth]{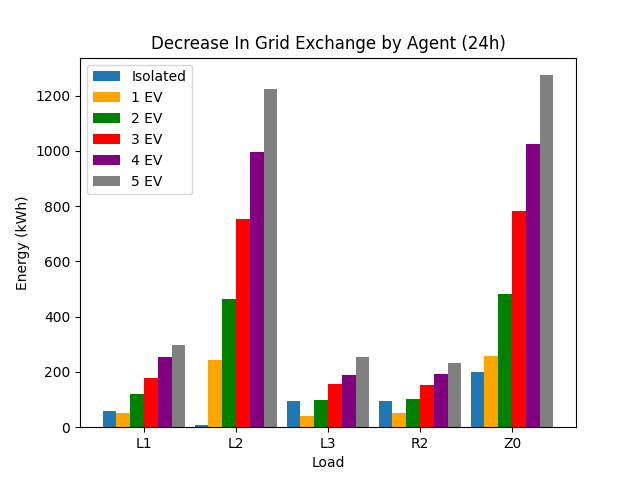}
    \caption{Energy traded within our EMS for each agent in each trading scenario, using our Greedy matching algorithm.}
    \label{Scale}
    \vspace{-0.9em}
\end{figure}

Fig. \ref{Scale} demonstrates the advantages of enabling EVs to interact with multiple Prosumers through a secure trading platform, as implemented in this research. Using the ``Isolated Grid Exchange" trading scenario from Table \ref{v2hBenchmark} and represented by ``Isolated" in Fig. \ref{Scale}, we establish a reference point for evaluating the benefits of allowing multiple connections and interactions between trading EVs and Prosumers. Once again this benefit is measured as the amount of energy in kWh agents were able to trade within the system instead of having to trade this energy with the main grid.

The benefits of allowing multiple EVs to interact with multiple Prosumers are immediately apparent, with one EV trading with five Prosumers (the ``1 EV" trading scenario) performing comparably to the ``Isolated" benchmark scenario. Across all agents, an aggregate 497.78\% decrease in energy exchange with the main grid was experienced in the ``1 EV" trading scenario. This benefit of decreased energy exchange with the main grid is skewed towards Agents ``L1" and ``Z0" which are the only agents to experience a net benefit in the ``1 EV" over the ``isolated" scenario. As we only accept one matching per per period, the energy previously stored for agents ``L1", ``L3", and ``R2" gets used by other agents, mainly ``L2" which is consuming energy during their energy production. As we increase the participating EVs in the grid, the benefit starts to become available to all agents. This occurs in the ``2 EV" trading scenario, all agents experience beneficial energy exchange, though this is minimal for ``L3" and ``R2" with an increase of 4.6\% and 9.9\% respectively.

We also observed roughly consistent increasing returns as the number of EVs increases. For example, if we take the aggregate \% increase of all agents from Fig. \ref{Scale}, as we increase the number of EVs from 1-2EVs, we get an increase in return of 657\%, and when we increase from 4-5EVs, we get an increase in return of 665\%.

Table \ref{lsapgreedy} shows the benefit of utilising matching algorithms enabled by load forecasting by demonstrating a performance comparison between the Greedy and Hungarian algorithms for each agent in different trading scenarios. The Hungarian algorithm outperformed the Greedy algorithm in all scenarios, increasing energy traded within our system between 5.4-27.8\%. The most significant differences occurred when the number of Prosumers greatly outnumbered the number of EVs with ``1 EV" and ``2 EV" trading scenarios experiencing the most significant benefit of 19.9\% and 27.8\% increase in energy trading within the developed system.

\begin{center}
    \begin{table}[h]
    \vspace{-1em}
    \centering
    \caption{Table showing the improvements of using Hungarian matching algorithm over the Greedy matching algorithm in terms of decreasing energy exchange with the main grid.}
    \begin{tabular}{|C{0.04\textwidth}|c|c|c|c|c|C{0.00001\textwidth}|C{0.05\textwidth}|}
    \cline{1-6} \cline{8-8}
       \textbf{No. EVs}  & \textbf{L1}                             & \textbf{L2}                              & \textbf{L3}                             & \textbf{R2}                             & \textbf{Z0}                             &  & \textbf{Agg. \% inc.}        \\ \cline{1-6} \cline{8-8} 
    1 EV & \cellcolor[HTML]{82FF82}49.2\% & \cellcolor[HTML]{FFEFEF}-6.1\%  & \cellcolor[HTML]{0AFF0A}96.2\% & \cellcolor[HTML]{61FF61}62.0\% & \cellcolor[HTML]{D1FFD1}18.1\% &  & \cellcolor[HTML]{CCFFCC}19.9 \\ \cline{1-6} \cline{8-8} 
    2 EV & \cellcolor[HTML]{8CFF8C}44.9\% & \cellcolor[HTML]{F9FFF9}2.5\%   & \cellcolor[HTML]{0FFF0F}94.1\% & \cellcolor[HTML]{43FF43}73.8\% & \cellcolor[HTML]{C0FFC0}24.5\% &  & \cellcolor[HTML]{B8FFB8}27.8 \\ \cline{1-6} \cline{8-8} 
    3 EV & \cellcolor[HTML]{B5FFB5}29.1\% & \cellcolor[HTML]{FFDFDF}-12.4\% & \cellcolor[HTML]{11FF11}93.2   & \cellcolor[HTML]{A5FFA5}35.4\% & \cellcolor[HTML]{E4FFE4}10.7\% &  & \cellcolor[HTML]{E1FFE1}12.0 \\ \cline{1-6} \cline{8-8} 
    4 EV & \cellcolor[HTML]{3EFF3E}75.6\% & \cellcolor[HTML]{FFE9E9}-8.7\%  & \cellcolor[HTML]{A2FFA2}36.5\% & \cellcolor[HTML]{D1FFD1}17.9\% & \cellcolor[HTML]{EFFFEF}6.4\%  &  & \cellcolor[HTML]{E5FFE5}10.3 \\ \cline{1-6} \cline{8-8} 
    5 EV & \cellcolor[HTML]{FFF8F8}-2.9\% & \cellcolor[HTML]{DCFFDC}13.9\%  & \cellcolor[HTML]{FFF8F8}-2.6\% & \cellcolor[HTML]{D6FFD6}16.1\% & \cellcolor[HTML]{FFFCFC}-1.2\% &  & \cellcolor[HTML]{F1FFF1}5.4  \\ \cline{1-6} \cline{8-8} 
    \end{tabular}
    \label{lsapgreedy}
    \end{table}
    \vspace{-2em}
\end{center}

\begin{figure}[h]
    \vspace{-1em}
    \centering
    \includegraphics[width=0.48\textwidth]{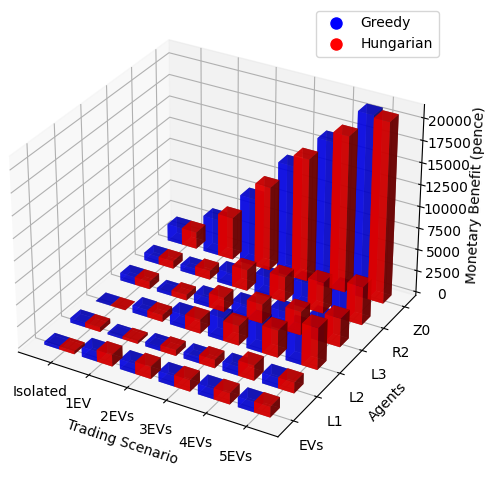}
    \caption{Monetary benefit for participating agents using each algorithm, implementing the suggested pricing behaviours.}
    \label{monetary}
\end{figure}

Another important factor of the proposed solution is monetary incentives, encouraging participation in the network. Using the static pricing behaviours, we have calculated the monetary benefit of trading in our system for each agent in each scenario shown in Fig. \ref{monetary}. 

The monetary benefit results indicate that all entities within our system experience a measurable benefit over trading with the main grid. The monetary benefit for EVs is calculated as profit and shown as an average per EV. The results show that each EV made between £3.39 and £12.84 for the 24-hour trading period in the Hungarian Algorithm. Notably, the maximum benefit was experienced when trading with 4 EVs while the minimum benefit was observed in the ``Isolated" Scenario. These findings indicate the potential for proposed solution to provide significant financial benefits to EV owners.

Furthermore, we also calculated the monetary value for Prosumers, which was the benefit of trading energy compared to trading all their energy with the main grid. However, This benefit was not equally distributed among all agents. In particular, the most beneficial entities were those that traded the most energy with the grid, with ``L2" and ``Z0" experiencing monetary benefits between £0.28-£48.08 and £19.63-£207.58, respectively. Moreover, all Prosumers in the proposed solution experienced maximum benefit in the ``5 EV" trading scenario, while experiencing minimum benefit in the ``Isolated" Trading Scenario. Interestingly, ``Z0" experienced the most significant benefit throughout all scenarios, highlighting the importance of considering the energy production profile of Prosumers when evaluating the financial benefits of our system.

Table \ref{fedVaryleave} shows the performance of the federated implementation in terms of energy utility, compared to using perfect energy predictions. These federated energy predictions assume that all 5 loads had an EV predicting their energy production and consumption in all trading periods using our federated LSTM forecasting.

The differences from using the federated learning predictions over the full predictions is minimal. In the case of ``Isolated" trading, the federated predictions perform as well as the perfect predictions with a 0\% change. The federated learning system performs similarly to a system with perfect energy predictions, only experiencing between -3.7\% and -1.5e-14\% energy traded within the system. However, there is an outlier in the data where in the case of 3 EVs, the federated prediction outperforms our perfect prediction model by 1.4\%.  

\vspace{-1.0em}
\begin{center}
\begin{table}[ht]
\caption{Performance comparison of federated predictions against perfect predictions using the Hungarian algorithm.}
\label{fedVaryleave}
\centering
\begin{tabular}{|@{}c|c@{}|}
\hline
\textbf{Trading Scenario} & \textbf{Federated \% change in Energy Utility} \\ \hline
Isolated         & 0.0\%                                      \\ \hline
1 EV             & -1.5e-14\%                                 \\ \hline
2 EV             & \cellcolor[HTML]{FFF5F5}-3.7\%             \\ \hline
3 EV             & \cellcolor[HTML]{FAFFFA}1.40\%             \\ \hline
4 EV             & \cellcolor[HTML]{FFFBFB}-1.7\%             \\ \hline
5 EV             & \cellcolor[HTML]{FFFCFC}-1.3\%             \\ \hline
\end{tabular}
\end{table}
\end{center}
\vspace{-2.5em}

\subsection{Discussion}
The secure management system we have developed enables the trading of energy between multiple EVs and Prosumers, leading to significant advantages such as decreased dependence on the main grid and monetary benefits for all participants. 
The analysis of results shows that the benefit to Prosumers in our simulation heavily depends on their load profiles. In terms of energy utility and decreased reliance on the main grid, the two most significant factors were the total amount of energy exchange and net energy production. Prosumers with a larger magnitude of energy exchange with the main grid experienced far greater absolute benefits than Prosumers with smaller energy exchanges with the main grid. The other most significant factor came in net energy production, where net energy producers experienced greater pricing benefits. The fact that our proposed system offers both financial and energy utility benefits highlights its feasibility as a promising option for managing DRER. 

However, this evidence can be circumstantial. Regarding pricing behaviours, we have assumed that our prices $P_{gb}$ and $P_{gs}$ are static. In contrast, energy prices are usually dynamic and can change throughout even a 24-hour period which could have some consideration in our results. Additionally, factors such as distribution losses and limited charging and discharging rates are not discussed. This could lead to bottlenecks in the system, reducing the displayed benefits. 

Furthermore, the EVs and Prosumers have clearly defined roles on opposite sides of the market allowing us to use LSAP. As such we never consider the potential of interactions between agents on the same side of the market. By allowing EVs and agents to trade energy with each other, we could extend our system to allow actively charging EVs to participate in the network charging from energy stored in EVs, and our Prosumers could trade without using an EV as a middleman. This would, however, decrease the benefit to the EVs, the main resource providers, and decreasing their benefit could reduce their incentive to participate.  

The finding that the percentage increase in benefit from adding EVs to the network is roughly consistent in Fig. \ref{Scale} has necessary implications for EV networks' optimal design and operation. By only allowing one matching per EV and Prosumer per period, we could limit the system in terms of scalability. While our results suggest that there is unutilised benefit to be gained from further increasing the number of EVs in the system, the rate of increase in benefits may eventually slow as energy resources from Prosumers become more limited. Therefore, the optimal design and operation of EV networks require a nuanced approach that considers factors such as the availability of energy resources, charging infrastructure, and demand patterns.

Furthermore, as the number of EVs and Prosumers approached equilibrium, the benefit of Hungarian over the Greedy algorithm in terms of energy utility decreased. When there are more Prosumers than EVs, Hungarian can exploit efficient matching allocations to a greater extent.

Finally, our LSTM predictions were shown to be successful by providing comparable results to our system with perfect energy prediction. This is because the resource capabilities of the EVs limit energy management. Therefore EVs can only partially accommodate all production and consumption within an energy period, and therefore the potential loss through inaccurate energy predictions is minimal. The worst-performing scenario experienced a decrease of 3.7\% for energy trading with a Hungarian algorithm and 3.9\% for prediction with the Greedy algorithm. If we were to scale this smart grid system to accommodate more EVs, we could see that the inaccurate federated learning predictions have a more significant effect on the system and may cause this disparity to increase.

In the case of 3 EVs, the forecasted predictions outperformed the perfect predictions by 1.40\%, this is likely due to the nature of the Hungarian matching algorithm which maximises energy exchange with matching periods. For example, if there is a miss-prediction on energy consumption, the Hungarian algorithm may prefer a matching that stores energy over a matching that releases energy. This would increase the amount of energy able to be released later in periods of consumption and increase the overall utility of the system. This raises an interesting direction for future research into time-dependent matching algorithms. By utilising forecasted energy predictions over longer periods, we could potentially implement more efficient and responsive matching algorithms that maximise energy utility.

\section{Concluding Remarks}
\label{sec:conclusions}
In this paper, a holistic DLT-based EMS is proposed that utilises EV energy storage capabilities in conjunction with using their computational abilities for load forecasting. This EMS allowed for efficient matching algorithms to maximise energy utility whilst providing measurable monetary benefits to all participating agents. We demonstrated that by allowing the interactions and trading between multiple EVs and Prosumers, we can increase both gross and net benefit to the agents. Finally, we identify some possible areas for future research, with more powerful load forecasting and time-dependent matching algorithms, different market mechanisms allowing homogeneous trading between agents, and more in-depth analysis of pricing behaviours.

\bibliographystyle{IEEEtran}
\bibliography{references.bib}

\begin{thebibliography}{10}
\providecommand{\url}[1]{#1}
\csname url@samestyle\endcsname
\providecommand{\newblock}{\relax}
\providecommand{\bibinfo}[2]{#2}
\providecommand{\BIBentrySTDinterwordspacing}{\spaceskip=0pt\relax}
\providecommand{\BIBentryALTinterwordstretchfactor}{4}
\providecommand{\BIBentryALTinterwordspacing}{\spaceskip=\fontdimen2\font plus
\BIBentryALTinterwordstretchfactor\fontdimen3\font minus \fontdimen4\font\relax}
\providecommand{\BIBforeignlanguage}[2]{{%
\expandafter\ifx\csname l@#1\endcsname\relax
\typeout{** WARNING: IEEEtran.bst: No hyphenation pattern has been}%
\typeout{** loaded for the language `#1'. Using the pattern for}%
\typeout{** the default language instead.}%
\else
\language=\csname l@#1\endcsname
\fi
#2}}
\providecommand{\BIBdecl}{\relax}
\BIBdecl

\bibitem{ZHANG20181}
\BIBentryALTinterwordspacing
C.~Zhang \emph{et~al.}, ``{Peer-to-Peer energy trading in a Microgrid},'' \emph{Applied Energy}, vol. 220, pp. 1--12, 2018. [Online]. Available: \url{https://www.sciencedirect.com/science/article/pii/S0306261918303398}
\BIBentrySTDinterwordspacing

\bibitem{IOTAP2P}
C.~Mullaney \emph{et~al.}, ``{Peer-to-Peer Energy Trading meets IOTA: Toward a Scalable, Low-Cost, and Efficient Trading System},'' in \emph{IEEE/ACM 15th International Conference on Utility and Cloud Computing (UCC)}, 2022, pp. 399--406.

\bibitem{Safdar2013AMicro-grids}
S.~Safdar \emph{et~al.}, ``{A survey on communication infrastructure for micro-grids},'' in \emph{2013 9th International Wireless Communications and Mobile Computing Conference, IWCMC 2013}, 2013, pp. 545--550.

\bibitem{kiaee2015estimation}
M.~Kiaee, A.~Cruden, and S.~Sharkh, ``Estimation of cost savings from participation of electric vehicles in vehicle to grid (v2g) schemes,'' \emph{Journal of Modern Power Systems and Clean Energy}, vol.~3, no.~2, pp. 249--258, 2015.

\bibitem{hu2017survey}
P.~Hu \emph{et~al.}, ``Survey on fog computing: architecture, key technologies, applications and open issues,'' \emph{Journal of network and computer applications}, vol.~98, pp. 27--42, 2017.

\bibitem{ouramdane2022home}
O.~Ouramdane \emph{et~al.}, ``Home energy management considering renewable resources, energy storage, and an electric vehicle as a backup,'' \emph{Energies}, vol.~15, no.~8, p. 2830, 2022.

\bibitem{parikh2019security}
S.~Parikh \emph{et~al.}, ``Security and privacy issues in cloud, fog and edge computing,'' \emph{Procedia Computer Science}, vol. 160, pp. 734--739, 2019.

\bibitem{ravi2022utilization}
S.~S. Ravi and M.~Aziz, ``Utilization of electric vehicles for vehicle-to-grid services: Progress and perspectives,'' \emph{Energies}, vol.~15, no.~2, p. 589, 2022.

\bibitem{vadi2019review}
S.~Vadi \emph{et~al.}, ``A review on communication standards and charging topologies of v2g and v2h operation strategies,'' \emph{Energies}, vol.~12, no.~19, p. 3748, 2019.

\bibitem{real-timeEMMicro}
Z.~Shen \emph{et~al.}, ``Real-time energy management for microgrid with ev station and chp generation,'' \emph{IEEE Transactions on Network Science and Engineering}, vol.~8, no.~2, pp. 1492--1501, 2021.

\bibitem{xu2022short}
J.~Xu and Y.~Huang, ``The short-term optimal resource allocation approach for electric vehicles and v2g service stations,'' \emph{Applied Energy}, vol. 319, p. 119200, 2022.

\bibitem{yi2015survey}
S.~Yi, C.~Li, and Q.~Li, ``A survey of fog computing: concepts, applications and issues,'' in \emph{Proceedings of the 2015 workshop on mobile big data}, 2015, pp. 37--42.

\bibitem{jaiswal2021fog}
R.~Jaiswal, R.~Davidrajuh, and S.~Wondimagegnehu, ``Fog computing for efficient predictive analysis in smart grids,'' in \emph{Proceedings of the International Conference on Artificial Intelligence and its Applications}, 2021, pp. 1--6.

\bibitem{zhang2022federated}
G.~Zhang, S.~Zhu, and X.~Bai, ``Federated learning-based multi-energy load forecasting method using cnn-attention-lstm model,'' \emph{Sustainability}, vol.~14, no.~19, p. 12843, 2022.

\bibitem{hou2016vehicular}
X.~Hou \emph{et~al.}, ``Vehicular fog computing: A viewpoint of vehicles as the infrastructures,'' \emph{IEEE Transactions on Vehicular Technology}, vol.~65, no.~6, pp. 3860--3873, 2016.

\bibitem{ning2019vehicular}
Z.~Ning, J.~Huang, and X.~Wang, ``Vehicular fog computing: Enabling real-time traffic management for smart cities,'' \emph{IEEE Wireless Communications}, vol.~26, no.~1, pp. 87--93, 2019.

\bibitem{Weiss2019AGenerators}
G.~Weiss, F.~D{\"{o}}rfler, and Y.~Levron, ``{A stability theorem for networks containing synchronous generators},'' \emph{Systems and Control Letters}, vol. 134, 2019.

\bibitem{melton2013gridwise}
R.~B. Melton, ``Gridwise transactive energy framework (draft version),'' Pacific Northwest National Lab.(PNNL), Richland, WA (United States), Tech. Rep., 2013.

\bibitem{wallnerstrom2018analysis}
C.~J. Wallnerstr{\"o}m and L.~B. Tjernberg, ``Analysis of the future power systems's ability to enable sustainable energy—using the case system of smart grid gotland,'' in \emph{Application of Smart Grid Technologies}.\hskip 1em plus 0.5em minus 0.4em\relax Elsevier, 2018, pp. 373--393.

\bibitem{Smart_Grids}
V.~C. Gungor \emph{et~al.}, ``A survey on smart grid potential applications and communication requirements,'' \emph{IEEE Transactions on Industrial Informatics}, vol.~9, no.~1, pp. 28--42, 2013.

\bibitem{Wang2019}
\BIBentryALTinterwordspacing
N.~Wang \emph{et~al.}, ``When energy trading meets blockchain in electrical power system: The state of the art,'' \emph{Applied Sciences 2019, Vol. 9, Page 1561}, vol.~9, p. 1561, 4 2019. [Online]. Available: \url{https://www.mdpi.com/2076-3417/9/8/1561/htm https://www.mdpi.com/2076-3417/9/8/1561}
\BIBentrySTDinterwordspacing

\bibitem{Shuaib2018}
K.~Shuaib \emph{et~al.}, ``Using blockchains to secure distributed energy exchange,'' \emph{2018 5th International Conference on Control, Decision and Information Technologies, CoDIT 2018}, pp. 622--627, 6 2018.

\bibitem{kong2021secure}
M.~Kong \emph{et~al.}, ``Secure and efficient computing resource management in blockchain-based vehicular fog computing,'' \emph{China Communications}, vol.~18, no.~4, pp. 115--125, 2021.

\bibitem{iqbal2021secure}
A.~Iqbal \emph{et~al.}, ``A secure and decentralized blockchain based ev energy trading model using smart contract in v2g network,'' \emph{IEEE Access}, vol.~9, pp. 75\,761--75\,777, 2021.

\bibitem{wen2020blockchain}
Y.~Wen \emph{et~al.}, ``Blockchain consensus mechanisms and their applications in iot: A literature survey,'' in \emph{Algorithms and Architectures for Parallel Processing: 20th International Conference, ICA3PP 2020, New York City, NY, USA, October 2--4, 2020, Proceedings, Part III 20}.\hskip 1em plus 0.5em minus 0.4em\relax Springer, 2020, pp. 564--579.

\bibitem{Popov2018TheTangle}
\BIBentryALTinterwordspacing
S.~Popov, ``{The tangle},'' \emph{ABA Journal}, no. FEB., pp. 1--25, 2018. [Online]. Available: \url{https://iota.org/\%0AIOTA Whitepaper.pdf}
\BIBentrySTDinterwordspacing

\bibitem{Salimitari2018ANetworks}
\BIBentryALTinterwordspacing
M.~Salimitari and M.~Chatterjee, ``{A Survey on Consensus Protocols in Blockchain for IoT Networks},'' 2018. [Online]. Available: \url{http://arxiv.org/abs/1809.05613}
\BIBentrySTDinterwordspacing

\bibitem{IOTAV2G}
V.~Hassija \emph{et~al.}, ``A blockchain-based framework for lightweight data sharing and energy trading in v2g network,'' \emph{IEEE Transactions on Vehicular Technology}, vol.~69, no.~6, pp. 5799--5812, 2020.

\bibitem{singla2018blockchain}
A.~Singla and E.~Bertino, ``Blockchain-based pki solutions for iot,'' in \emph{2018 IEEE 4th international conference on collaboration and internet computing (CIC)}.\hskip 1em plus 0.5em minus 0.4em\relax IEEE, 2018, pp. 9--15.

\bibitem{jia2018double}
B.~Jia \emph{et~al.}, ``Double-matching resource allocation strategy in fog computing networks based on cost efficiency,'' \emph{Journal of Communications and Networks}, vol.~20, no.~3, pp. 237--246, 2018.

\bibitem{mills2007dynamic}
G.~A. Mills-Tettey, A.~Stentz, and M.~B. Dias, ``The dynamic hungarian algorithm for the assignment problem with changing costs,'' \emph{Robotics Institute, Pittsburgh, PA, Tech. Rep. CMU-RI-TR-07-27}, 2007.

\bibitem{Gomes2019-xv}
L.~Gomes and Z.~Vale, ``{uGIM}: week monitorization data of a microgrid with five agents (04/08/2019 - 10/08/2019),'' 2019.

\bibitem{gaete2021open}
C.~Gaete-Morales \emph{et~al.}, ``An open tool for creating battery-electric vehicle time series from empirical data, emobpy,'' \emph{Scientific data}, vol.~8, no.~1, p. 152, 2021.

\bibitem{kostopoulos2020real}
E.~D. Kostopoulos, G.~C. Spyropoulos, and J.~K. Kaldellis, ``Real-world study for the optimal charging of electric vehicles,'' \emph{Energy Reports}, vol.~6, pp. 418--426, 2020.

\bibitem{GridBuyPrice}
\BIBentryALTinterwordspacing
``{Energy Cost UK}.'' [Online]. Available: \url{https://www.nimblefins.co.uk/average-cost-electricity-kwh-uk}
\BIBentrySTDinterwordspacing

\bibitem{GridSellPrice}
\BIBentryALTinterwordspacing
``{Energy Cost UK}.'' [Online]. Available: \url{https://www.britishgas.co.uk/help-and-support/my-account/smart-export-guarantee}
\BIBentrySTDinterwordspacing

\bibitem{iotawiki_2023}
\BIBentryALTinterwordspacing
``Iota wiki,'' Feb 2023. [Online]. Available: \url{https://wiki.iota.org/develop/welcome}
\BIBentrySTDinterwordspacing

\bibitem{Pytorch}
\BIBentryALTinterwordspacing
``{Pyorch LSTM Documentation}.'' [Online]. Available: \url{https://pytorch.org/docs/stable/generated/torch.nn.LSTM.html}
\BIBentrySTDinterwordspacing

\end{thebibliography}
\end{document}